# Fingerprint of vortex-like flux closure in isotropic Nd-Fe-B bulk magnet

Mathias Bersweiler,[1] Yojiro Oba,[2] Evelyn Pratami Sinaga[1], Inma Peral[1], Ivan Titov[1], Michael P. Adams[1], Venus Rai[1], Konstantin L. Metlov[3] and Andreas Michels[1]

[1]*Department of Physics and Materials Science, University of Luxembourg, 162A Avenue de la Faïencerie, L-1511 Luxembourg, Grand Duchy of Luxembourg*
[2]*Materials Sciences Research Center, Japan Atomic Energy Agency, 2-4 Shirakata, Tokai, Ibaraki, 319-1195, Japan*
[3]*Institute for Numerical Mathematics RAS, 8 Gubkina Street, Moscow, GSP-1 119991, Russian Federation*

Corresponding author: andreas.michels@uni.lu

**Abstract** – Taking advantage of recent progress in neutron instrumentation and in the understanding of magnetic-field-dependent small-angle neutron scattering, here, we study the three-dimensional magnetization distribution within an isotropic Nd-Fe-B bulk magnet. The magnetic neutron scattering cross section of this system features the so-called spike anisotropy, which points towards the presence of a strong magnetodipolar interaction. This experimental result combined with a damped oscillatory behavior of the corresponding correlation function and recent micromagnetic simulation results on spherical nanoparticles suggest an interpretation of the neutron data in terms of vortex-like flux-closure patterns. The field-dependent correlation length $L_c$ is well reproduced by a phenomenological power-law model. While the experimental neutron data for $L_c$ are described by an exponent close to unity ($p$ = 0.86), the simulation results yield $p$ = 1.70, posing a challenge to theory to include vortex-vortex interaction effects.

## I. Introduction

Permanent magnets are defined by their high remanent magnetization and high coercivity. Ideally, their magnetization in the remanent state should be as uniform as possible within the bulk. Yet, because of the high coercivity (related to the large magnetic anisotropy) saturating them is not an option for quantifying how far their remanent state is away from the uniform one and what kind of magnetization nonuniformities develop at remanence.

These nonuniformities are usually micrometer in size and their direct observation within the bulk of the magnet is only possible with tomographic techniques. The current state of the art in x-ray magnetic nanotomography [1–3], which utilizes the magnetic circular dichroism effect, is strongly tied to the details of the absorption edge of a particular element (e.g., Gd) and is not applicable to an arbitrary magnetic material without significant adjustments. By contrast, magnetic small-angle neutron scattering (SANS) is universally applicable to any kind of magnetic material, and can disclose the magnetic microstructure in the bulk and on the relevant mesoscopic length scale of ∼1–1000 nm [4–6].

Sintered Nd-Fe-B is nowadays one of the most used permanent magnets that finds application in many key industry sectors [7]. Commercial-grade magnets consist of highly magnetic (and with high crystalline magnetic anisotropy) $Nd_2Fe_{14}B$ grains, sintered and magnetized in such a way that their average magnetization points (mostly) towards the same direction. However, due to the magnetostatic interaction the grains tend to develop various flux-closure magnetization textures, which, depending on the grain size and other technological parameters, result in a multitude of multidomain structures (observable e.g. by surface microscopy and/or Bitter pattern techniques [8,9]), lowering the energy product of the magnet. This problem becomes less pronounced for smaller grains (especially the ones that are embedded in the bulk of the magnet), whose magnetization becomes almost uniform, but even then the magnetostatic energy favors a subtle magnetization curling [10]. Also, the surface flux-closure domains are almost always very different from the magnetic texture in the bulk, which for Nd-Fe-B still remains unobserved, but (for macroscopic sample sizes) makes the major contribution towards its remanent magnetization.

As shown by Vivas *et al.* [11] using micromagnetic simulations, the flux-closure structures in a set of spherical magnetic nanoparticles can be analyzed using the correlation function analysis of the corresponding magnetic SANS cross section. In this study, we apply this specific analysis technique to real experimental scattering data and find strong evidence for the existence of vortex-like flux-closure textures within the grains of a Nd-Fe-B magnet.

**II. Experimental Details**

For the neutron experiments, a circular-shaped disk of a commercially available sintered isotropic (i.e., untextured) Nd-Fe-B permanent magnet (original anisotropic grade: N52) with a diameter of 22.0 mm and a thickness of 420 µm was prepared. A summary of the microstructural and magnetic characterization results of the Nd-Fe-B specimen can be found in the Supplemental Material [to be inserted] (see also Ref. [12]). The neutron experiments were conducted at the instrument SANS-J at the JRR-3 research reactor in Tokai, Japan [13]. Figure 1(a) sketches the scattering geometry used in this work. The neutron experiments were done at room temperature using an unpolarized neutron beam with a mean wavelength of $\lambda = 6.5$ Å and a wavelength broadening of 14 % (FWHM). By employing a focusing neutron-lens setup, the accessible magnitude of the momentum-transfer vector $q$ ranged between about 0.003 nm$^{-1}$ ≤ $q$ ≤ 0.3 nm$^{-1}$, so that real-space structures on a scale of a few nm up to a few µm could be probed. It is this particular feature of the neutron instrumentation at SANS-J that allows us to access large-scale magnetization fluctuations, well beyond the capabilities of conventional SANS instruments (see Ref. [13] for further details). A magnetic field $\boldsymbol{H}_0$ was applied perpendicular to the incident neutron beam ($\boldsymbol{H}_0 \parallel \boldsymbol{e}_z \perp \boldsymbol{k}_0$). Neutron data were recorded by reducing the magnetic field from 10 T (maximum field available) down to 0 T. The neutron-data reduction (corrections for background scattering and sample transmission) was performed using an in-house program written in Igor Pro software (WaveMetrics). For the neutron-data analysis, the experimental purely magnetic SANS cross sections $d\Sigma_{\mathrm{mag}}/d\Omega$ were determined by subtracting the total (nuclear + magnetic) SANS cross section $d\Sigma/d\Omega$ measured at 10 T (approach-to-saturation regime) from the ones measured at lower fields. This subtraction procedure eliminates the (field-independent) nuclear scattering contribution, and results in a purely magnetic difference SANS cross section. The method has been successfully used in several other studies, e.g., to investigate the magnetization profile within nanoparticles [14], or to disclose the magnetic microstructure in off-stoichiometric Heusler alloys [15].

## III. Magnetic correlation function and correlation length

The quantity of interest in our experiments is the magnetic correlation function $C(r)$, which provides information on the real-space correlations of the three-dimensional magnetization vector field on the mesoscopic length scale [6]. We have numerically computed the $C(r)$ from the experimental purely magnetic SANS cross section $d\Sigma_{mag}/d\Omega$ via an indirect Fourier transform (IFT) technique, based on the following Fourier transform:

$$C(r) = \frac{1}{r} \int_0^\infty \frac{d\Sigma_{mag}}{d\Omega}(q) \sin(qr) q \, dq \quad . \quad (1)$$

The numerical inversion method was introduced in the 1970s by Glatter [16]; for the particular case of magnetic SANS, it represents a fast and robust means to obtain model-free information that reflects the real-space magnetization of magnetic materials (e.g., nanoparticles [17,18] or bulk ferromagnets [15,19]); see Ref. [20] for technical information and a discussion of the IFT approach. We have also computed the magnetic correlation length $L_c$, which characterizes the average distance over which fluctuations of the magnetization vector field are correlated. There are several methods discussed in the literature to define and quantify $L_c$; e.g., $L_c$ can be determined from the logarithmic derivative of the magnetic correlation function $C(r)$ in the limit $r \to 0$ (Ref. [21]), or it can be defined as the value of $r$ for which $C(r) = C(0)\exp(-1)$ (Ref. [22]). These two methods for estimating $L_c$ focus on the behavior of $C(r)$ at small distances $r$. Here, in order to obtain the full information on the vortex structures (over the entire $r$ range), we determined $L_c$ at a particular external field $H_0$ according to

$$L_c(H_0) = \frac{\int_0^\infty r \, C(r, H_0) dr}{\int_0^\infty C(r, H_0) dr} \quad . \quad (2)$$

We emphasize that for exponentially-decaying correlations all three definitions yield the same correlation length. In our analysis, we also display results for the so-called distance distribution function $P(r)$, which is related to the correlation function via $P(r) = r^2 C(r)$. Due to the $r^2$ factor, features at medium and large distances $r$ are more pronounced in $P(r)$ than in $C(r)$. Note that the error bars of the one-dimensional magnetic SANS data (shown in Fig. 2 below) are taken into account in the computation of the $C(r)$ and $P(r)$ curves (Fig. 3). The here plotted $C(r)$ and $P(r)$ correspond to the results with the highest evidences [20].

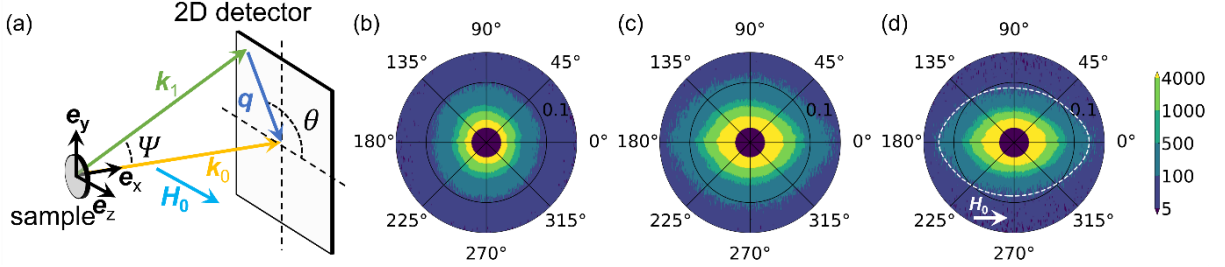

FIG. 1. (a) Sketch of the scattering geometry used for the magnetic SANS experiments. The momentum-transfer vector $q$ corresponds to the difference between the wavevectors of the incident ($k_0$) and the scattered ($k_1$) neutrons, i.e., $q = k_0 - k_1$. The magnetic field $H_0$ is applied perpendicular to the incident neutron beam, i.e., $H_0 \parallel e_z \perp k_0$. For small-angle scattering (i.e., $\Psi \ll 1$), the component $q_x$ of $q$ is smaller than the other two components $q_y$ and $q_z$, so that only correlations in the plane perpendicular to the incident neutron beam are probed. (b) and (c) Experimental two-dimensional total (nuclear + magnetic) unpolarized SANS cross section $d\Sigma/d\Omega$ of isotropic Nd-Fe-B permanent magnet at the selected fields of 10 T [(b), near saturation] and 0 T [(c), remanence]. (d) Corresponding purely magnetic SANS cross section $d\Sigma_{mag}/d\Omega$ obtained by subtracting (b) from (c). Note that in (b) to (d) the SANS data are plotted in polar coordinates with $q$ in nm$^{-1}$, $\theta$ in degrees, and the intensity in cm$^{-1}$. White dashed line in (d): guide to the eyes to emphasize the spike-type angular anisotropy due to the magnetodipolar interaction.

**IV. Results and Discussion**

Figures 1(b) and (c) display typical examples of the experimental two-dimensional (2D) total SANS cross section $d\Sigma/d\Omega$ of isotropic Nd-Fe-B at the selected fields of 10 T (near saturation) and 0 T (remanence), respectively. As can be seen, near saturation the pattern is slightly elongated perpendicular to the magnetic-field direction. This feature in $d\Sigma/d\Omega$ is the signature of the so-called "$\sin^2(\theta)$-type" angular anisotropy due to predominantly longitudinal magnetization fluctuations. By contrast, at remanence, one can clearly observe the emergence of maxima in $d\Sigma/d\Omega$ along the magnetic-field direction. This observation indicates the built-up of a more complex magnetization texture. Figure 1(d) shows the corresponding 2D purely magnetic SANS cross section $d\Sigma_{mag}/d\Omega$ obtained from the subtraction of (b) from (c). In this way, the sharp maxima along the field direction become more clearly visible, thereby revealing the so-called "spike-type" angular anisotropy. As detailed by Périgo *et al.* [23], this particular feature in $d\Sigma_{mag}/d\Omega$ is a consequence of the magnetostatic pole-avoidance principle; it is due to a nonzero magnetic volume charge density of the magnetization that shows up as a characteristic angular dependence of the Fourier modes of the magnetostatic field, which in turn determine the Fourier components of the magnetization and, therefore, of $d\Sigma_{mag}/d\Omega$.

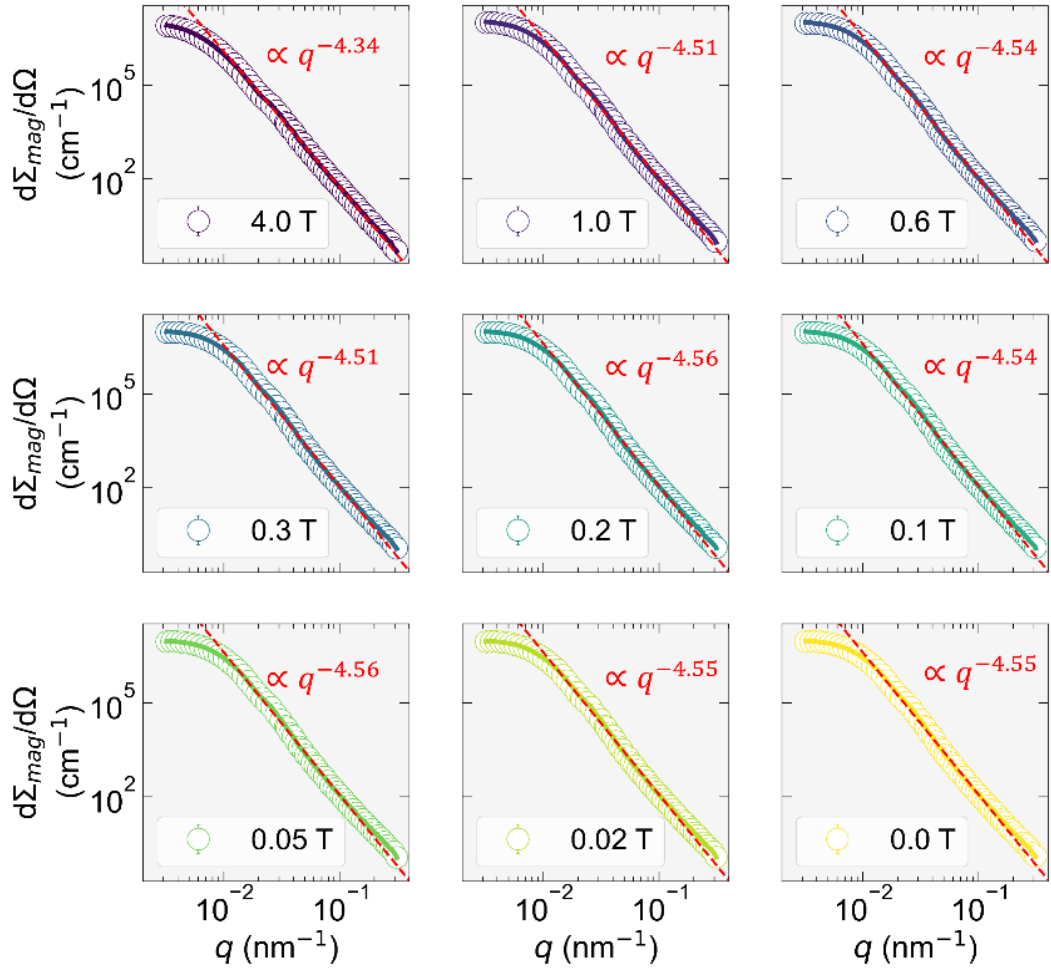

FIG. 2. Magnetic-field dependence of the (over $2\pi$) azimuthally-averaged purely magnetic SANS cross section $d\Sigma_{mag}/d\Omega$ of isotropic Nd-Fe-B permanent magnet (log-log scales). Error bars of $d\Sigma_{mag}/d\Omega$ are smaller than the symbol size. Red dashed lines: Extrapolation of $d\Sigma_{mag}/d\Omega \propto q^{-n \pm 0.05}$. These Porod fits were restricted to $0.034 \leq q \leq 0.14$ nm$^{-1}$. Colored solid lines: Reconstructed $d\Sigma_{mag}/d\Omega$ based on the indirect Fourier transform (IFT) of the numerically computed $P(r)$ shown in Fig. 3.

Figure 2 presents the (over $2\pi$) azimuthally-averaged purely magnetic SANS cross sections $d\Sigma_{mag}/d\Omega$ of isotropic Nd-Fe-B permanent magnet. In the low-$q$ region, the $d\Sigma_{mag}/d\Omega$ curves reveal a Guinier-type behavior around $q \sim 0.006$ nm$^{-1}$, indicating that the probed magnetic microstructure is at least 1 μm in size. In the high-$q$ region, the $d\Sigma_{mag}/d\Omega$ exhibit a $q^{-n}$-type decay, where the asymptotic power-law exponent $n$ is found to be consistently larger than the value of $n = 4$, corresponding to scattering from particles with sharp interfaces or from exponentially-correlated magnetization fluctuations. The finding of values $n > 4$ supports the notion of dominant spin-misalignment scattering in isotropic Nd-Fe-B permanent magnet, and is consistent with theoretical predictions and experimental results [6,12,24]. This observation is a consequence of the fact that magnetic SANS has its origin in smoothly-varying continuous magnetization profiles, rather than in sharp discontinuous magnetic-moment variations.

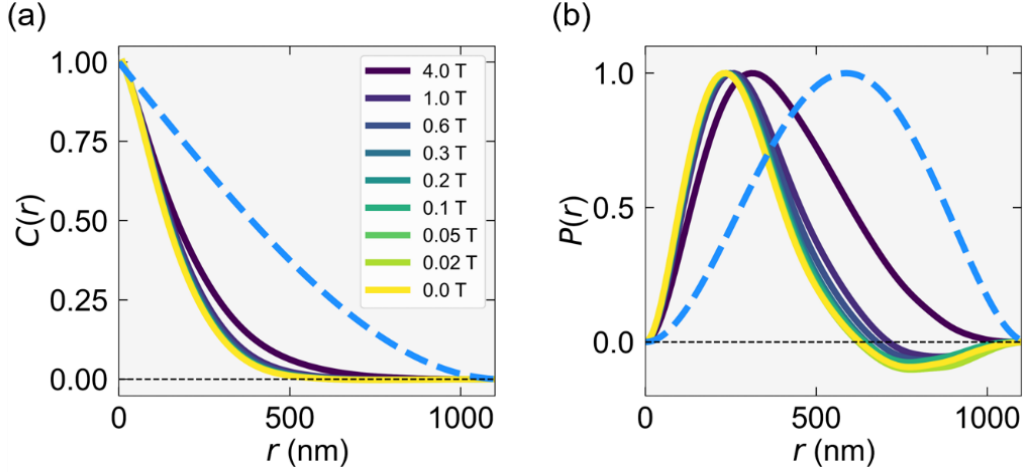

FIG. 3. (a) Field dependence of the magnetic correlation function $C(r)$ [Eq. (1)], which was numerically computed via an indirect Fourier transform (IFT) of the experimental $d\Sigma_{mag}/d\Omega$ data shown in Fig. 2. Blue dashed lines: (a) $C(r) = 1 - 3r/(4R) + r^3/(16R^3)$ and (b) $P(r) = r^2 C(r)$ for a uniformly magnetized sphere with a radius of $R = 560$ nm. The latter value is estimated by extrapolating Eq. (3) to a saturating field of 20 T and using $L_C^{H\to\infty} = \frac{8}{15}R$. Note that the $C(r)$ at each field have been normalized by their respective maximum values. (b) Corresponding (normalized) distance distribution functions $P(r)$.

Figure 3 shows the magnetic-field dependence of the magnetic correlation function $C(r)$ and of the corresponding distance distribution function $P(r)$ computed via IFT. In the following, we focus the discussion on the behavior of the $P(r)$ [rather than on the $C(r)$], since due to the $r^2$ factor features at medium and large distances are more pronounced in $P(r)$ than in $C(r)$. Depending on the magnetic field strength, two distinct behaviors are observed for the $P(r)$. At the highest field, $P(r)$ exhibits a deformed bell-like shape, which is reminiscent of an approximately spherical correlation volume. Compared to the case of a uniformly magnetized sphere (blue dashed lines in Fig. 3), it is reasonable to assume that this $P(r)$ corresponds to a magnetic structure with a size of about 1 μm and with an internal spin configuration that deviates only slightly from the perfect alignment along the field direction. By reducing the magnetic field strength, the $P(r)$ disclose a "damped oscillatory" behavior with negative values, and a zero-crossing shifting to smaller $r$. As previously discussed by Vivas et al. [11], the combination of these two features can be used as a strong indication for the presence of an inhomogeneous magnetization texture; more specifically, numerical micromagnetic computations revealed that these features in $P(r)$ are related to the emergence of a vortex-like flux closure in nanoparticles.

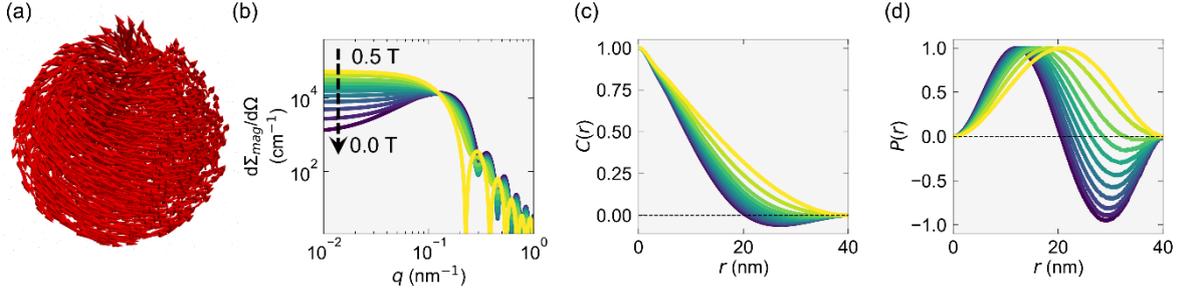

FIG. 4. Signature of a vortex-like magnetization texture in the magnetic SANS observables. (a) Snapshot of a vortex-type magnetization distribution (at remanence) in a 40-nm-sized spherical nanoparticle, obtained using the open-source software code MuMax3 [25]. For the micromagnetic simulations of the real-space magnetization distribution, the sphere volume was discretized into cubic cells with a size of $2 \times 2 \times 2$ nm$^3$, and the magnetic material parameters of Fe were used [26,27] (b) Magnetic-field dependence of the (over $2\pi$) azimuthally-averaged magnetic SANS cross section $d\Sigma_{mag}/d\Omega$, numerically computed from the simulated real-space magnetization distribution at selected fields $H_0$ [ranging from 0.5 T (near saturation) to remanence with an increment of 0.05 T]. (c) Corresponding magnetic correlation functions $C(r)$ [Eq. (1)]. (d) Magnetic distance distribution functions $P(r) = r^2 C(r)$.

Figure 4 displays micromagnetic simulation results for the specific case of a vortex-like magnetization texture in a 40-nm-sized spherical nanoparticle; details on the micromagnetic simulations using MuMax3 [25] can be found in Ref. [27]. Due to the large magnetocrystalline anisotropy of Nd-Fe-B, the corresponding single-domain size (for a spherical particle) is relatively large, about 210 nm [26]. To eventually disclose a vortex structure in a Nd-Fe-B sphere requires a much larger particle diameter of the order of 500 nm, which from the micromagnetic simulation point of view is computationally very challenging. Therefore, to see the general field-dependent behavior of the correlation length of a single vortex, we assumed for the simulation results displayed in Fig. 4 the materials parameters of Fe, for which a vortex-type structure already appears for diameters larger than about 20 nm [11,27]. These results are expected to be qualitatively transferable to the case of Nd-Fe-B. The value of 40 nm for the Fe sphere diameter is indeed arbitrary, and we could have equally well chosen a larger size. The simulations on the 40 nm sphere serve as a guide that also allow us to test the predictions for the correlation length [Eq. (2)] in the limiting case of a saturated sphere (see Fig. 5 below).

The most characteristic signature of a vortex-type magnetization distribution [as shown in Fig. 4(a)] in the neutron-scattering data is a damped oscillatory behavior of the corresponding correlation functions $C(r)$ and $P(r)$ with a shift of the zero crossing to smaller $r$ with decreasing field [compare Fig. 4(c) and 4(d)]. The oscillatory behavior, which is preserved in the presence of a particle-size distribution [27], can be readily explained as follows: a vortex structure is characterized by relatively large spin variations, circulating about the vortex axis, so that the autocorrelation of a vortex with its displaced "ghost" gets dominated (at some particular value of displacement $r$) by "anticorrelations" with negative values in both $C(r)$ and $P(r)$. We would like to emphasize that the emergence of a vortex-type spin structure and the concomitant oscillatory feature in the $P(r)$ is a direct consequence of the dipolar interaction [11,27], which is also of decisive importance for the appearance of the spike-type pattern in the two-dimensional experimental $d\Sigma_{mag}/d\Omega$ [see Fig. 1(d)]. In fact, as demonstrated in Ref. [23], without the dipolar interaction the spike feature is absent in $d\Sigma_{mag}/d\Omega$. Moreover, while the simulation results at low fields reveal the absence of a Guinier behavior in $d\Sigma_{mag}/d\Omega$ at the smallest momentum transfers [compare Fig. 4(b)], this is not seen in our experimental neutron data [Fig. 2], which at all fields studied feature a Guinier-type behavior followed by a plateau region. In this regard we emphasize that the presence of a vortex-type texture in the sample does not imply a peak-type dependence of the one-dimensional $d\Sigma_{mag}/d\Omega$; even a plateau at small $q$ is compatible with the existence of vortex structures [compare Fig. 4(b) and 4(d)].

An anticorrelation with a corresponding zero crossing in $C(r)$ and $P(r)$ is generally expected whenever (at some distance $r$) an antiparallel spin component (relative to the spin at the origin) appears. An example is the spin structure in the vicinity of a pore (magnetic hole) or of a second-phase particle, where the jump in the saturation magnetization at the interface gives rise to a dipolar stray field that "imprints" its structure into the surrounding matrix phase [28]. However, for the present isotropic Nd-Fe-B sample, we have no evidence for the existence of nanosized pores or nanosized second-phase particles in the microstructure, and we do not observe the corresponding clover-leaf type angular anisotropy in the experimental neutron data (that such a dipole-field induced perturbation would produce).

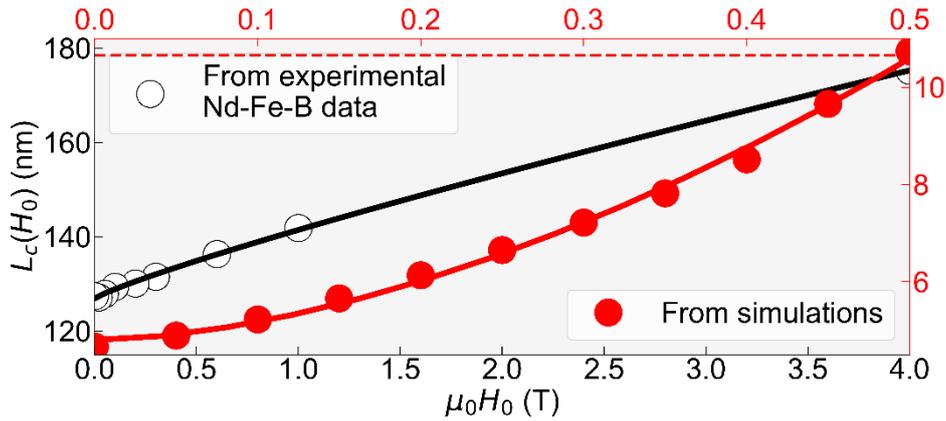

FIG. 5. (○) Field dependence of the magnetic correlation length $L_c$ obtained from the computed $C(r)$ data shown in Fig. 3(a) and using Eq. (2). (●) $L_c(H_0)$ obtained from the micromagnetic simulation data shown in Fig. 4(c) (note the different scales). Solid lines: fits to Eq. (3). The red dashed line represents the theoretical limit $L_c^{H \to \infty} = \frac{8}{15}R$ for a uniformly magnetized 40-nm-sized sphere using $C(r) = 1 - 3r/(4R) + r^3/(16R^3)$ and Eq. (2).

Figure 5 presents the field dependence of the magnetic correlation length $L_c(H_0)$ obtained from the experimental neutron scattering data of isotropic Nd-Fe-B and from the numerical micromagnetic simulations of a single 40-nm-sized spherical nanoparticle exhibiting a vortex-type magnetization distribution. In the case of the isotropic Nd-Fe-B sample (micromagnetic simulations on a single nanoparticle) $L_c$ increases from ~ 127 (4.7) nm at remanence to ~175 (10.3) nm at the highest field of 4 T (0.5 T). So far there exists no theoretical model that is able to describe the behavior of $L_c(H_0)$ for the case of a vortex-like flux-closure texture. An excellent description of the $L_c(H_0)$ data is obtained using the following power-law expression (solid lines in Fig. 5):

$$L_c(H_0) = L_c(H_0 = 0) + \beta H_0^p \quad , \qquad (3)$$

where $L_c(0) = 127.0 \pm 0.3$ nm, $\beta = 14.5 \pm 0.5$ nm/T$^p$, and $p = 0.86 \pm 0.02$ for the case of the isotropic Nd-Fe-B sample, and $L_c(0) = 4.8 \pm 0.1$ nm, $\beta = 19.0 \pm 1.0$ nm/T$^p$, and $p = 1.70 \pm 0.09$ for the simulation data [$H_0$ in Eq. (3) is in Tesla]. A model similar to Eq. (3) has already been used by Sonier *et al.* [29] to explain the field dependence of the magnetic penetration depth in the vortex state of a type-II superconductor. Certainly, the analogy to the superconductors seems to be rather far fetched, however, the functional form of Eq. (3) is, given the definition for the correlation length [Eq. (2)], plausible for vortex-type spin structures in spherical correlation volumes: at low fields, when the vortex is present, one expects an $L_c$ that is significantly smaller than the size of the sphere, while, towards saturation, $L_c$ increases with field and takes on its maximum value (here: $L_c^{H \to \infty} = \frac{8}{15} R$). This prediction, which is well followed in the simulations (see Fig. 5), also seems to describe the experimental data. However, to unambiguously demonstrate the suitability of Eq. (3), further neutron experiments at larger field values are necessary.

The observed difference in the power-law exponents between simulation ($p = 1.70$) and experiment ($p = 0.86$) might be related to the possible presence of vortex-vortex interactions in the Nd-Fe-B sample, which are absent in the simulations. A better comprehension of the magnetic field dependence of the magnetic correlation length $L_c$ requires the further extension of the micromagnetic SANS theory to include interacting vortex-like structures [30,31]. Finding the exponent that describes the field evolution of $L_c$ is an important open question in this regard.

The present unpolarized neutron measurements on a polycrystalline sample cannot discern a possible preferred sense of rotation of the vortices, e.g., when the applied field is reduced and vortex nucleation takes place with a vortex orientation that is aligned with the nearest easy-axis direction. In this respect, neutron polarization analysis might provide information on the presence of vortices with a net chirality.

## V. Conclusion

We have investigated the spin microstructure of an isotropic Nd-Fe-B bulk magnet using magnetic field-dependent small-angle neutron scattering (SANS) combined with micromagnetic simulations. Thanks to a focusing neutron-lens setup, we could access the long-range magnetic correlations in the magnetic SANS cross section $d\Sigma_{mag}/d\Omega$. The two-dimensional $d\Sigma_{mag}/d\Omega$ features a spike-type angular anisotropy, which is a consequence of the magnetodipolar interaction and the ensuing pole avoidance principle. Analysis of the magnetic correlation and distance distribution functions, numerically computed from the experimental $d\Sigma_{mag}/d\Omega$ via an indirect Fourier transform method, suggests the emergence of an internal vortex-like flux-closure magnetization distribution by reducing the magnetic field strength. The field dependence of the corresponding magnetic correlation length can be well described by a phenomenological power-law expression [Eq. (3)]. The investigation of vortex-like flux-closure patterns in isotropic Nd-Fe-B might be extended to commercial-grade Nd-Fe-B, which differs from the present isotropic sample only by the field-aided alignment step of the crystallites during the sintering procedure. For commercial-grade textured Nd-Fe-B, the appearance of vortices might reduce the remanent magnetization and, hence, limit the performance of the magnet. Finally, our study underlines that magnetic SANS combined with micromagnetic simulations is a promising approach towards the resolution of three-dimensional mesoscale spin structures in bulk materials.

**Acknowledgements**

Financial support by the National Research Fund of Luxembourg (AFR Grant No. 15639149 and PRIDE MASSENA Grant) and by KAKENHI (Grant No. 19K05102) is gratefully acknowledged. Konstantin

L. Metlov acknowledges the support of the Russian Science Foundation under Project No. RSF 21-11-00325. We thank the Japan Atomic Energy Agency for the provision of neutron beamtime at the SANS-J instrument and Dr. Shibabrata Nandi and Dr. Oleg Petracic (Jülich Centre for Neutron Science) for magnetization measurements.